\documentclass{aastex62}

\graphicspath{{./}{figures/}}

\begin{document}

\title{The shape of sunspots and solar activity cycles \footnote{Released on February, 14th, 2022}}

\correspondingauthor{Andrey Tlatov}
\email{tlatov@mail.ru}

\author[0000-0002-6286-3544]{Andrey G. Tlatov}
\affil{Kislovodsk Mountain astronomical station of the Pulkovo observatory \\
100 Gagarina str.,Kislovodsk, 357700, Russia}


\begin{abstract}

The paper presents the results of the analysis of the geometric characteristics of sunspots for the period of 19-24 cycles of activity. The shape of sunspots was studied on the basis of the method of normalization of images of sunspots to study the average profile of the spot. The deviation of the shape of sunspots from the axisymmetric configuration is investigated.
It was found that the spots, as a rule, have an ellipsoid shape, and the major axis of the ellipse has a predominant inclination to the equator, opposite in the Northern and Southern hemispheres.
The angle of inclination of the sunspot axis corresponds to the angle of inclination of the bipoles in the activity cycles. The relationship between the shape of sunspots in the current cycle and the amplitude of the next cycle of activity is found. The greater the elongation along the longitude
of the current cycle of spots, the higher the next cycle of activity will be.

\end{abstract}


\section{Introduction} \label{sec:intro}

Archival photographic images of the Solar photosphere according to various
observatories and modern observations allow us to study in detail the structure
and fine structure of sunspots as one of the most remarkable manifestations of solar activity. Ever since the first systematic observations of the Sun's disk, which can be attributed to the 18-th century, astronomers have sought not just to take into account the number of spots on a given day, but to make as detailed sketches of the shape of the spots as possible, thereby emphasizing the diversity of their manifestations \citep{Carrington}. Later, this knowledge formed the basis of the classification system of spots and spot groups, used in various modifications to this day. 

At the same time, a wide range of issues continues to be debatable. It is generally assumed that the sunspots formed at the places where strong magnetic fields enter the photosphere are located slightly below the outer boundary of the photosphere and represent certain depressions in its surface. This conclusion is based on both theoretical models of spot formation and projection effects observed when spots move across the Sun's disk. One of these effects is called the Wilson effect  and consists in changing the visible part of the spot when it is observed at different angles (in the center of the disk or near the limb)  \citep{Wilson65, Wilson68}. Further studies have shown that the manifestations of the Wilson effect have differences in the eastern and western hemispheres  \citep{Obashev,  Collados}.  To study the features of the shape of sunspots, it is useful to have an idea of how a spot looks on average and its typical parameters \citep{Bray}. 

In \citep{Illarionov}, a method was proposed for studying the average profile of a spot on a disk by scaling and superimposing images of all spots and constructing a histogram of the brightness distribution of the combined image. It was shown that the average shape of sunspots has a teardrop shape and is elongated in the longitude direction.
In this paper, we have analyzed the shapes of sunspots according to the processing of photoheliograms from the archive of the Kislovodsk Mountain Astronomical Station (KMAS) for the period from 1956 to 2021.

\section{Data and processing methods} \label{sec:data}

The Kislovodsk Mountain Astronomical Station has been performing daily observations of sunspots since 1948. During this time, a large archive of photographic plates (1948-2010) and digital images (2010- present) of the solar photosphere has been accumulated.

\begin{figure}[ht!]
\plotone{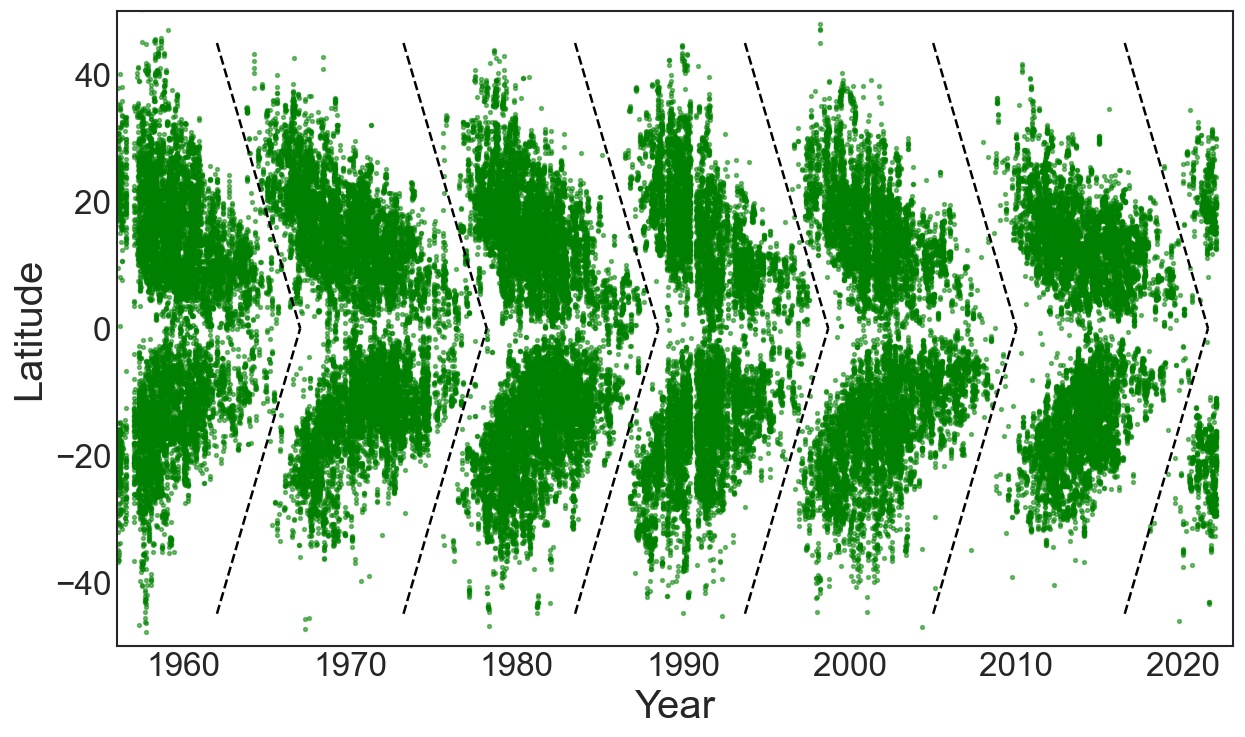}
\caption{Latitudinal-temporal diagram of the distribution of groups of sunspots. Dotted lines represent the sections of the cycle boundaries. \label{fig:fig1}}
\end{figure}

We digitized photographic plates, which made it possible to reconstruct the activity of sunspots using computer recognition methods \citep{Tlatov}.   Digitization data from 1956 are presented on the website \url{https://observethesun.ru}. 

During processing, we memorized not only tabular information, such as coordinates, area, group membership, but also information about the configuration of the shape of sunspots. This allows you to analyze the configurations of the shape of sunspots. Also, the representation of geometric information allows you to visually assess the level of solar activity, the relative location of solar activity, and their geo-efficiency in the case of flare and eruptive processes.

We used data on the characteristics of individual sunspots obtained by processing daily observations of KMAS photoheliograms in the period 1956-2021. In total, more than $77\cdot 10^3$ groups of spots were identified Figure \ref{fig:fig1}. In these, we selected $\sim 96 \cdot 10^3$  individual spots with an area of more than $S>30 \ \mu hm$. To analyze the shape, we considered the location of the photosphere-penumbra boundary.

\begin{figure}
\plottwo{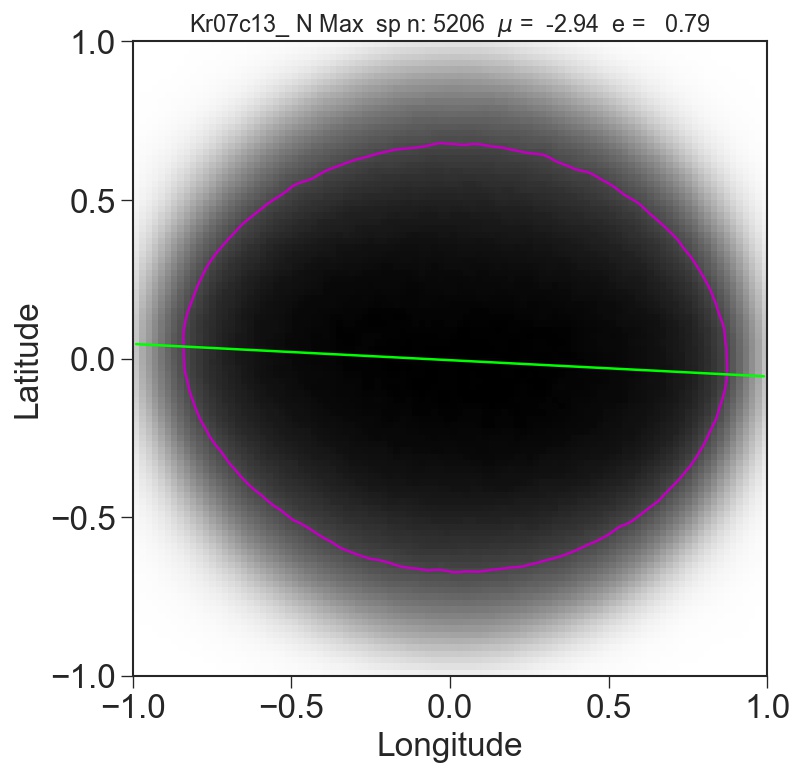}{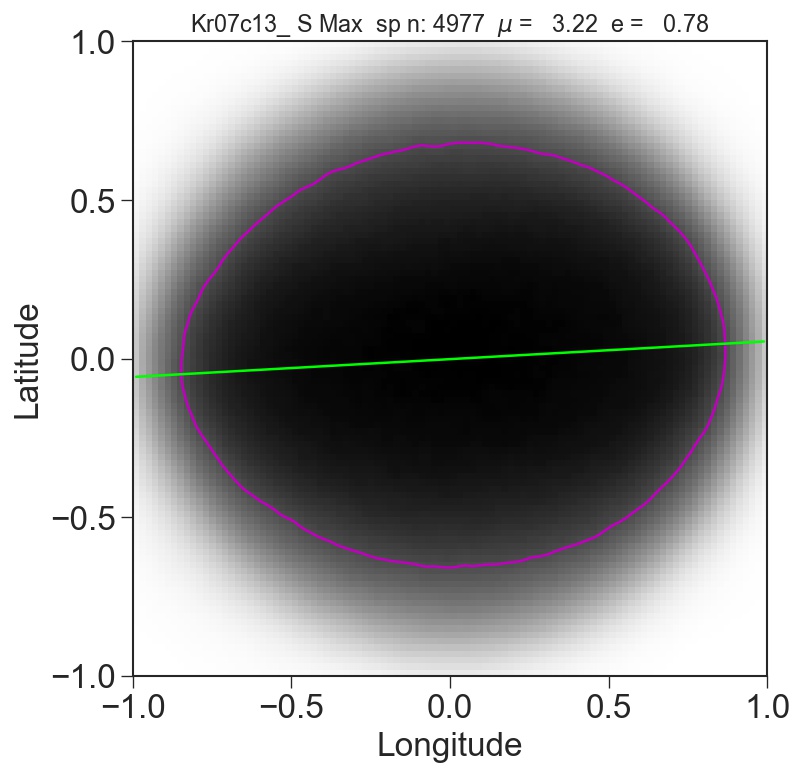}
\caption{The shape of sunspots in the Northern (left) and Southern (right) hemispheres.\label{fig:fig2}}
\end{figure}

To determine the direction of elongation, there are a number of approaches, from which the least squares method and the contour method can be distinguished. The least squares method searches for the direction along which all the points of the spot are located most densely, namely, have the smallest sum of squared distances \citep{Illarionov}. In the contour method used in this work, the asymmetry of density level  in the distribution density matrix is investigated.

\begin{figure} 
\plotone{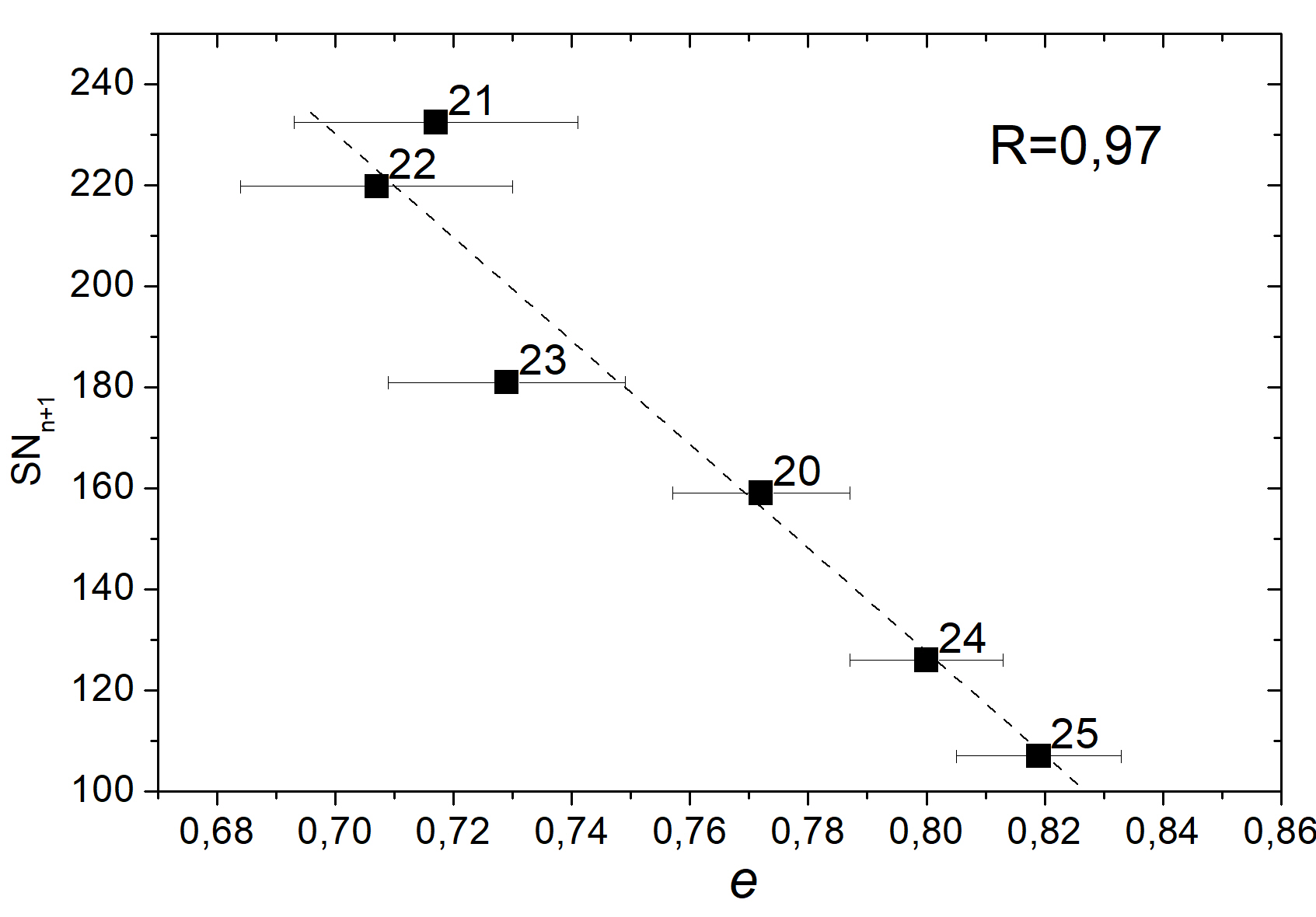}
\caption{The relationship of the solar activity cycle of the subsequent $SN_{n+1}$ activity cycle on the shape parameter of the sunspots $e$. The parameter e was calculated for the maximum spots in the group in terms of area. The error of determining the value of e is presented. \label{fig:fig3}}
\end{figure}

To exclude the influence of projection effects, the direction of the elongation of the spot is determined not in flat, but in spherical coordinates. To do this, we translated the pixels of the image into heliographic coordinates.  Then, similarly to \citep{Illarionov}, the image is scaled in such a way that the spot is written off into a single circle with the center coinciding with the center of the spot. All the resulting images are superimposed on each other, and a histogram of the brightness distribution is built based on these.

\begin{figure} 
\plotone{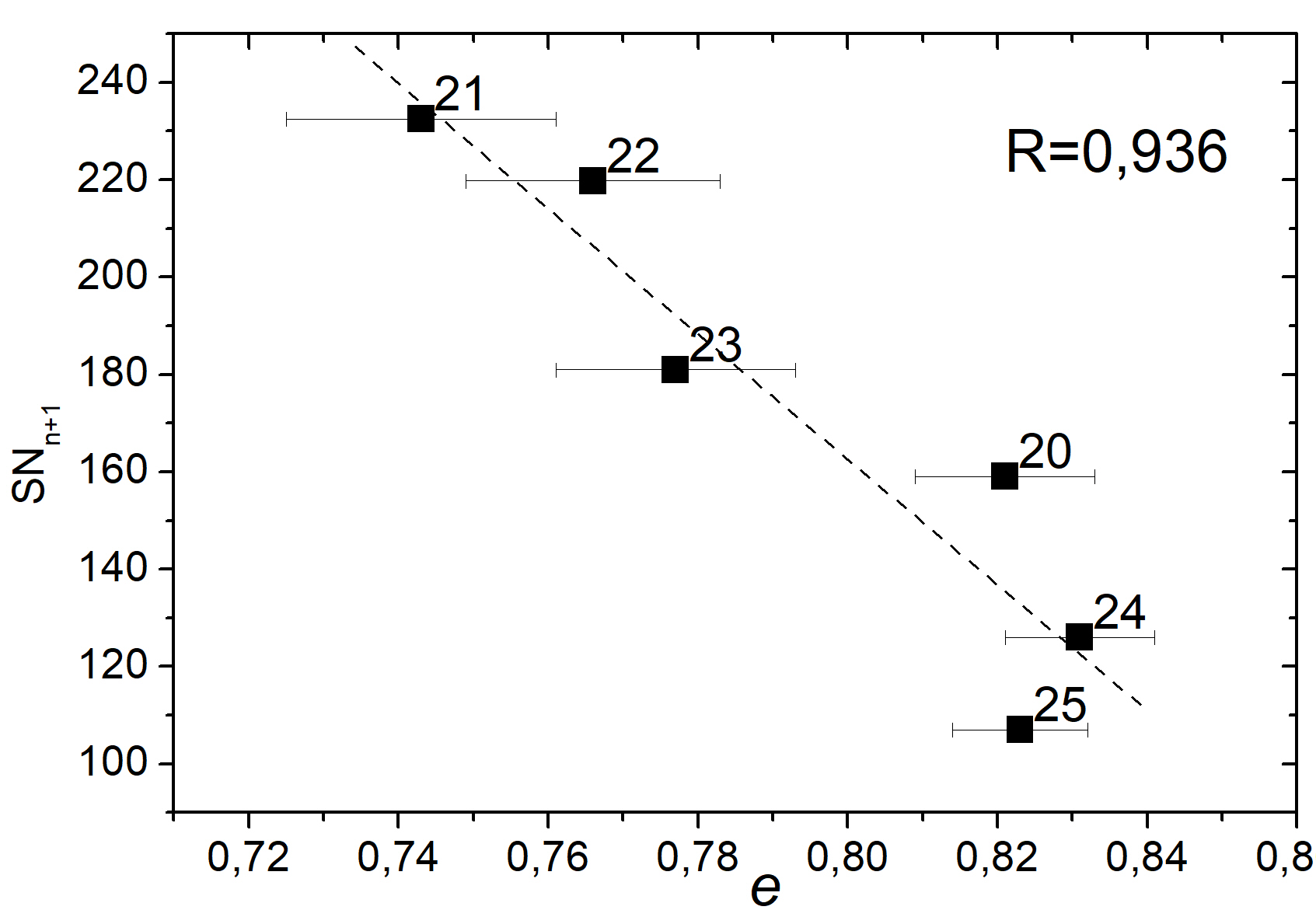}
\caption{The same as in Figure \ref{fig:fig3}, but for the second largest spots in groups. \label{fig:fig4}}
\end{figure}

Figure \ref{fig:fig2} shows the average spot profile for spots of maximum area in groups, for sampling spots with areas greater than $S>30 \ \mu hm$ from the central part of the disk ($r/R< 0.7$). In the period 1956-2021, according to KMAS data, we identified $\sim 5000$ spots in each hemisphere. The red line shows an ellipse inscribed in the distribution density for the value $k= 1.3\cdot Av$, where $Av$ is the average value of the distribution. The direction of the major axis of the ellipse is shown by the green line. The center of the spot corresponds to the position (0, 0).

\begin{figure}
\plottwo{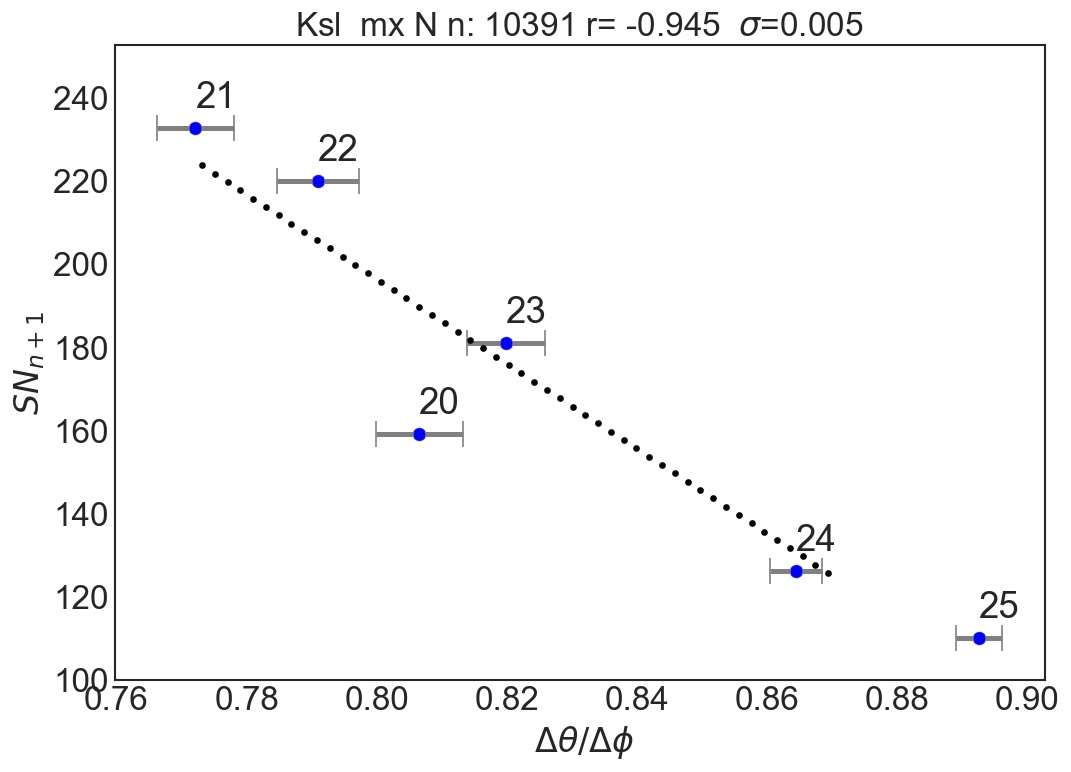}{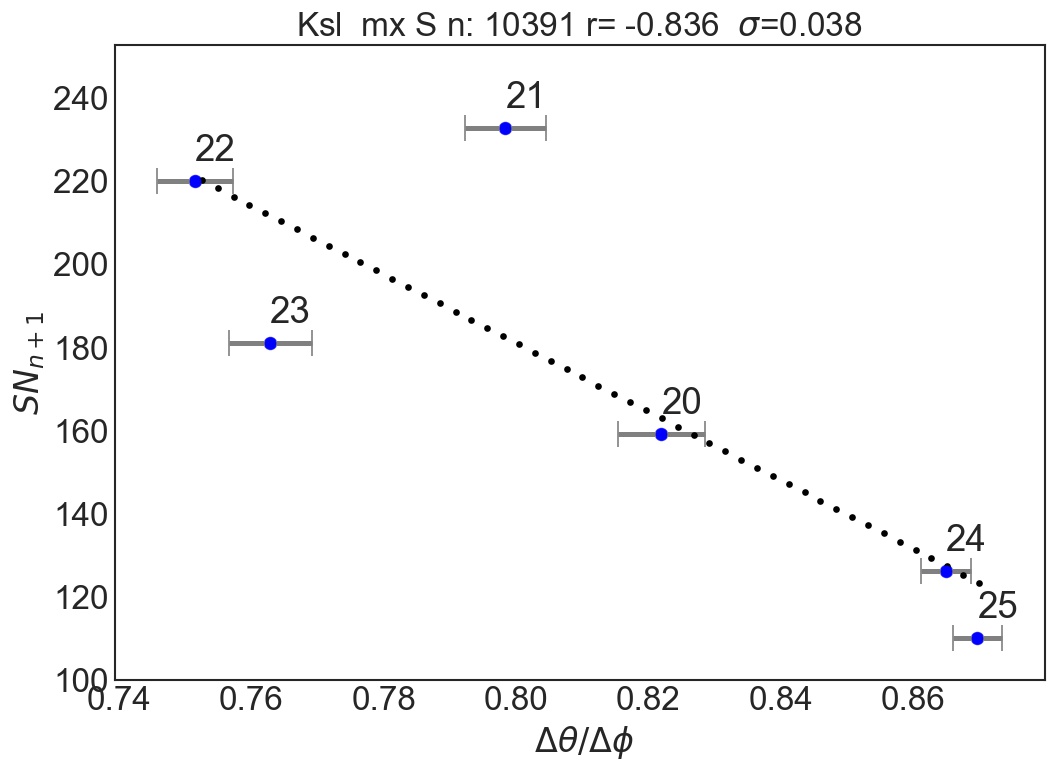}
\caption{The average ratio $\Delta  \theta / \Delta \phi $ calculated for the maximum cycles in the group and the amplitude of the next activity cycle.(Left) for the Northern Hemisphere. (Right) for the Southern Hemisphere.\label{fig:fig5}}
\end{figure}

The oblateness of spots can be estimated by a simpler method, for example, using the ratio of the size of the spots in latitude $\Delta \theta$ to the size in longitude $\Delta \phi$. Figure \ref{fig:fig5} shows the regression of the amplitudes of the $SN_{n+1}$ cycles from the ratio $\Delta \theta / \Delta \phi$ separately for the Northern and Southern hemispheres.  The correlation value turned out to be $r_{N}= 0.94$ and $r_{S}=0.83$ for the $N$ and $S$ hemispheres.

We can note that the major axis of the ellipse has an inclination with the solar equator, such that the eastern parts of the spots are further away from the equator than the western ones. Thus, for the Northern and Southern hemispheres, they have an inclination with respect to the equator in the same direction as the axes of the bipolar groups of sunspots. At the same time, we could not establish a clear change in the angle of inclination of the spots from latitude, similar to Joy's law \citep{Hale}.

Consider the parameter characterizing the elongation of the shape of the spot $e=b/a$, where $b$ and $a$ are the minor and major axes of the inscribed ellipse. Figure \ref{fig:fig3} shows the values of parameter e measured in cycles 19-24 and the amplitudes of the next activity cycle. The amplitudes of the activity cycles were taken for the new version of the spot index \citep{Usoskin}.  For cycle 25, we took the value $SN_{25} =110$, since this value lies on the regression line (Figure \ref{fig:fig3}).  The dependence of the amplitude SN on the value e for cycles n is expressed by $SN_{n+1}=948.01(\pm 102) - 1025.5(\pm 135)\cdot e_{n}$ at correlation $r=0.96$. The 25-th  cycle of activity currently lasts only two years. But according to these data, $e_{25}= 0.906(\pm 0.005)$, which corresponds to the activity level of $SN_{26} \sim 50$.

A significant relationship between the parameter of the shape of the spots e and the amplitude of the next cycle of $SN_{n+1}$ activity exists not only for the maximum spots in the group, but, for example, for the second for the area of the spots (Figure \ref{fig:fig4}). The correlation turned out to be less than for the maximum spots in the group $r=0.93$, but also quite high.

\section{Discussion} \label{sec:disc}

The performed analysis established new properties about the shape of sunspots. The method of averaging the shape of the spots over a long period showed that the sunspots differ from the round shape and are elongated along the longitude.  In this case, the spots have a slope with respect to the solar equator. The eastern parts of the spots are located farther from the equator than the western ones. This is similar to the slope of the bipoles of sunspots group.

In \citep{Illarionov}, when studying the differences in the visibility of the leading and tail spots in the Eastern and Western hemispheres, it was shown that the leading spots have an angle of inclination of $\sim 17^o$ with a vertical. The spots are formed when the magnetic field tubes exit onto the photosphere.  If the tubes have an angle of inclination with a vertical, even round tubes in cross-section on the photosphere will look like an ellipse.  Another factor that can influence the difference between the shape and the circle with the round shape of the tubes is the Wilson effect \citep{Wilson65, Wilson68}. When the spots are sunk under the photosphere near the limb, the spots will look elongated. It is possible that the elongated shape of the spots is also formed under the influence of additional forces, for example, the effect of matter flows on the magnetic field tubes. For example, the flow of meridional circulation.

However, the most likely mechanism for the formation of elliptical sunspots may be differential rotation in latitude. Indeed, in this case, the high-latitude parts of the spots will shift to the east, and the low-latitude ones to the west. In this case, the shape of the spots will stretch in a direction parallel to the equator, and the long axis of the spots will have a slope with respect to the equator. That is, to have a shape close to Figure \ref{fig:fig2}.

We carried out additional calculations of the dependence of parameter e on various parameters: latitude, distance from the center of the Sun's disk and on the size of the spots. The value of the parameter e depends on latitude and was $e \sim 0.83$ near the equator and $e \sim 0.74$ at latitudes $30^o$. At the same time, the dependence of the parameter e on the distance from the center of the Sun $r/R$, up to the values $r/R <0.7$, varies slightly. As the area of the spots increases, the parameter e decreases, i.e. large spots are more elongated than small ones.  These factors suggest that the elongated shape of the spots cannot be explained only by the Wilson effect.
It will be possible to assess the effect of the deviation of the force tubes from the vertical direction by analyzing the difference in the intensity of the magnetic field measured in the Eastern and Western hemispheres.



\end{document}